\documentclass[twocolumn,twoside]{article}
\usepackage{epsfig,overcite}
\usepackage{revmacs,reprintport}

\newcommand{\taumax}{\tau_{\rm max}}
\newcommand{\tauf}{\tau_{\!f}}
\newcommand{\taucorr}{\tau_{\rm corr}}
\newcommand{\Tg}{T_g}
\newcommand{\Tf}{T_f}
\newcommand{\Qts}{Q^{\ddag}}

\begin{document}

\newcommand{\addressbreak}{, }

\title{Diffusive Dynamics of the Reaction Coordinate for Protein
  Folding Funnels}

\author{N.~D.~Socci and
    J.~N.~Onuchic} 
\address{Department of Physics-0319\addressbreak
    University of California at San Diego\addressbreak
    La Jolla, California 92093}

\author{P.~G.~Wolynes}
\address{School of Chemical Sciences\addressbreak
    University of Illinois\addressbreak
    Urbana, Illinois 61801\\[.5\baselineskip]
    In press: {\em J.\ Chem.\ Phys.}, April 15, 1996}

\maketitle

\def\figureA{
\begin{figure}
\centerline{\epsfig{file=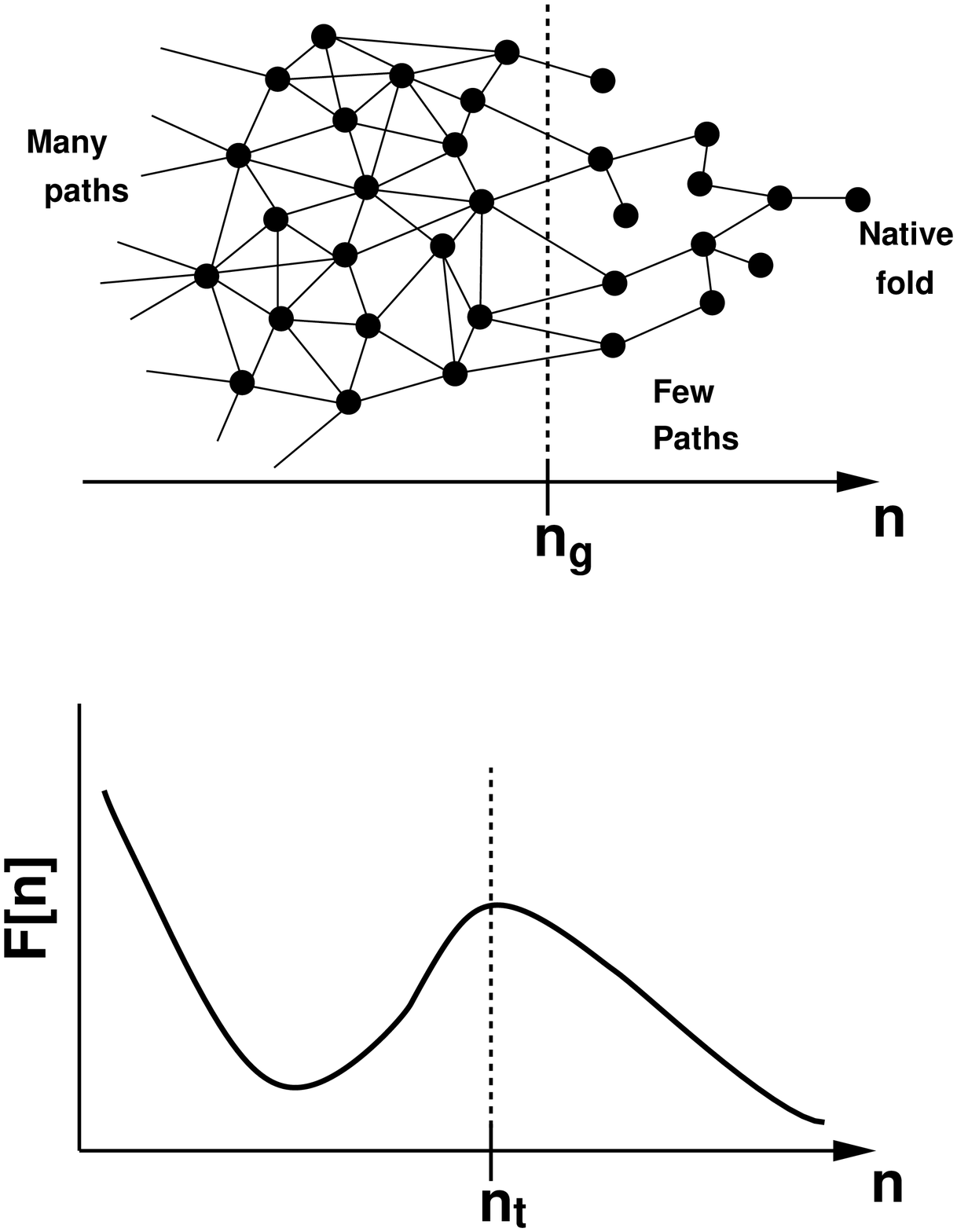,width=\figurewidth}}
\caption{The top figure shows the kinetic landscape versus an
  arbitrary reaction coordinate ($n$). The vertical line shows the glass
  transition point ($n_g$) in which the behavior changes from many
  paths to few paths. In addition to the native state there are a
  number of local minimum. The bottom figure shows the free energy
  plotted as a function of the same coordinate. The line at $n_t$
  denotes the transition region.}
\label{fig:landscape}
\end{figure}
}

\def\figureB{
\begin{figure}
\centerline{\epsfig{file=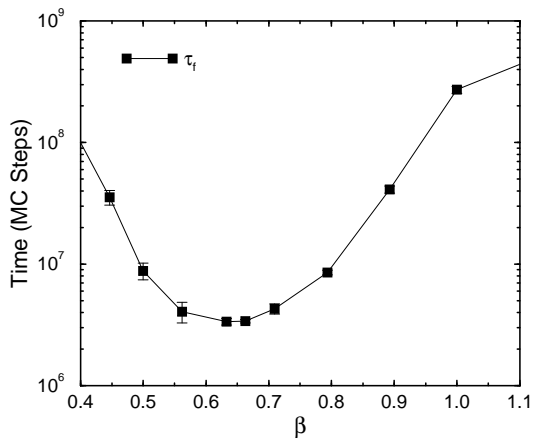,width=\figurewidth}}
\caption{Mean first passage time ($\tauf$) is plotted as a function of
  the inverse temperature ($\beta$). Note, parabolic dependence of the
  folding time on $\beta$.}
\label{fig:tauf}
\label{fig:parabola}
\end{figure}
}

\def\figureC{
\begin{figure}
\centerline{\epsfig{file=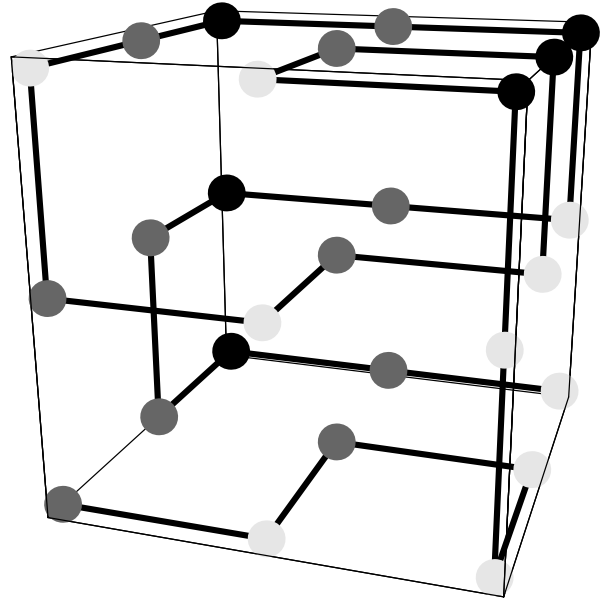,width=\figurewidth}}
\caption{The native state (minimum energy) cube for the sequence
  studied in this work. The sequence ({\tt
    ABABBBCBACBABABACACBACAACAB}) consist of three monomer types. }
\label{fig:cube}
\end{figure}
}

\def\figureD{
\begin{figure}
\centerline{\epsfig{file=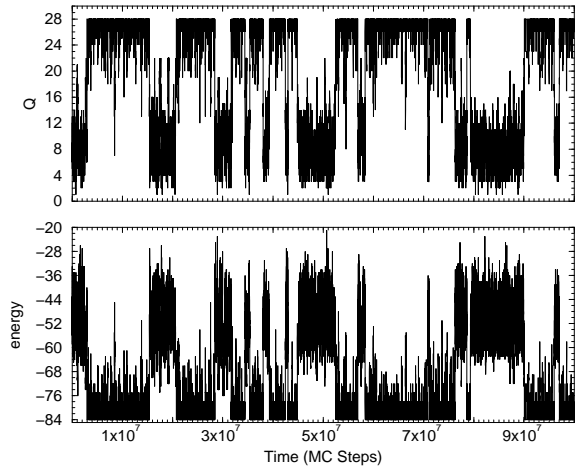,width=\figurewidth}}
\caption{The energy and order parameter ($Q$, the number of {\em
    correct} native contacts) plotted as a function of time for a
  sample folding simmulation. The temperture is the folding
  temperature ($\Tf=1.51$). The time is roughly 30 times the folding
  time ($\tauf\approx3\times10^6$). The plot shows the two-state-like
  behavior of this system with transition between the native state
  ($E=-84$, $Q=28$) and a molten globule region ($E\approx50$,
  $Q\approx=7$). The transition are extremly rapid with respect to the
  folding time. In addition there are significant fluctations about
  the two states.}
\label{fig:traj}
\end{figure}
}

\def\figureE{
\begin{figure*}
  \centerline{\epsfig{file=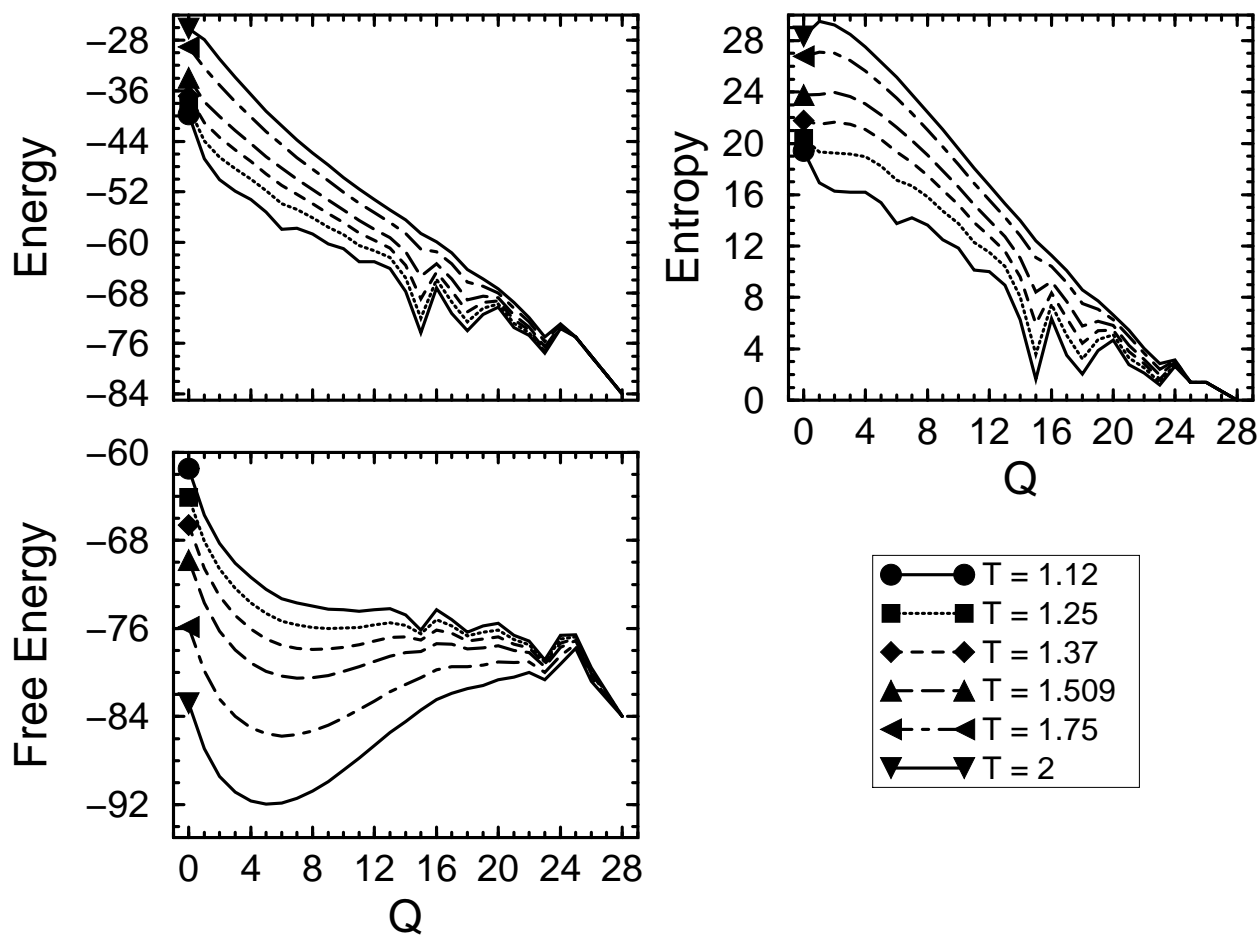,width=2\dblfigurewidth}}
\caption{Energy, entropy and free energy as a function of $Q$ for
  several temperatures calculated using the Monte Carlo histogram
  method~\protect\cite{Socci95}. For a broad range of temperatures the
  free energy has a clear double well form: one at the native state
  ($Q=28$) and one at the molten globule region ($Q\approx7$). The
  double well form is consistent with the two state behavior scene in
  figure~\protect\ref{fig:traj}. Above the glass transition
  temperature ($T_g\sim1$) there is a significant energy and entropy
  gradient between the molten globule region and the transition region
  ($Q\approx15$).  Both $\Delta E$ and $\Delta S$ are less than zero,
  so the unfavorable reduction in entropy is offset by a loss in
  energy, leaving a small free energy barrier at the folding
  temperature ($T=1.509$). As the temperature approaches the glass
  temperature the energy and entropy gradients decrease as does the
  free energy barrier. However the folding time diverges due to the
  increase in the diffusion constant of the system (i.e., the
  roughness of the energy landscape).}
\label{fig:freeq}
\end{figure*}
}

\def\figureF{
\begin{figure}
  \centerline{\epsfig{file=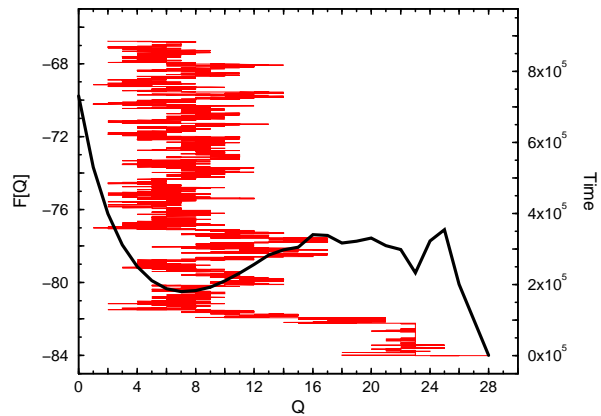,width=\figurewidth}}
\caption{A folding trajectory ($Q(\tau)$) superimposed on the free
  energy ($F[Q]$) at the folding temperature. The trajectory is
  approximately $9\times10^5$ Monte Carlo steps long (approximately
  27\% of $\tauf$). The right axis shows the time counted backwards
  from the folded state. Note, most of the time is spent diffusing
  in the molten globule region. Once the barrier is surmounted folding
  proceeds rapidly. Numerous recrossings of the barrier region
  indicate the need to use a diffusive theory for this reaction
  coordinate note transition state theory.}
\label{fig:trajfq}
\end{figure}
}

\def\figureG{
\begin{figure}
  \centerline{\epsfig{file=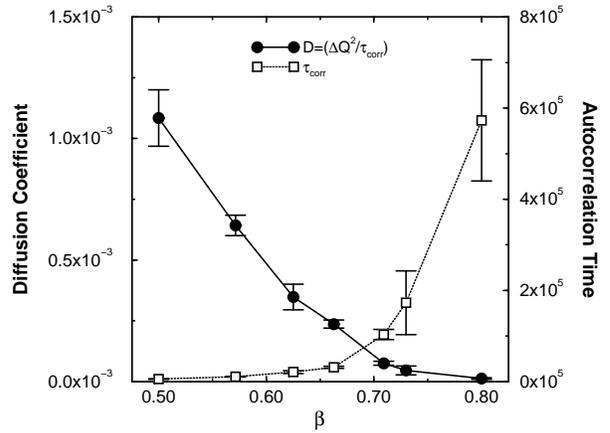,width=\figurewidth}}
\caption{Autocorrelation time ($\taucorr$) and diffusion coefficient
  ($\Delta Q^2/\taucorr$) plotted as a function of the inverse
  temperature ($\beta$). At low temperature (high $\beta$) the
  diffusion coefficient decreases rapidly while the correlation time
  increases, indicating a slow down of the dynamics due to trapping in
  local minimum.}
\label{fig:tcorr}
\end{figure}
}

\def\figureH{
\begin{figure}
  \centerline{\epsfig{file=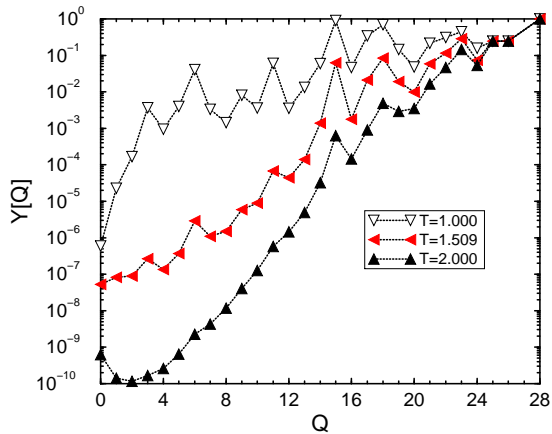,width=\figurewidth}}
\caption{Qualitative replica symmetry breaking value $Y(Q)= \sum_i
  \left[P_{i}(Q)\right]^{2}$ plot at three temperatures ($\Tg$, $\Tf$
    and $2\Tg$). Note that although at the global glass temperature
    ($\Tg$) discrete states are apparent even for small degrees of
    nativeness, at the folding temperature the discrete intermediates
    are largely native-like.}
\label{fig:yq}
\end{figure}
}

\def\figureI{
\begin{figure}
\centerline{\epsfig{file=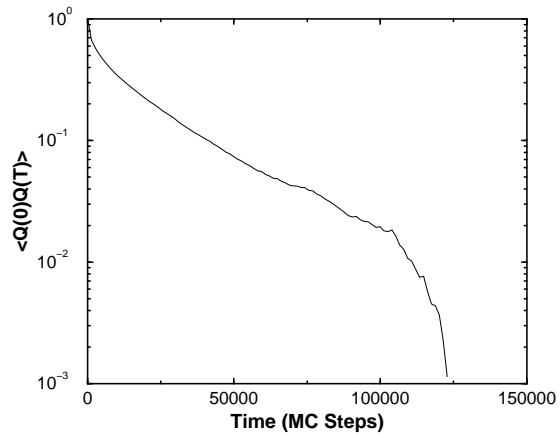,width=\figurewidth}}
\caption{Autocorrelation function for $Q$ (see eq.~\protect\ref{eq:autocorr})
  plotted on a semi-log scale. $T=\Tf$. The correlations have an
  initial rapid decay followed by a slower decay. By fitting a sum of
  exponentials (usually 3 or 4) the autocorrelation time can be
  determined.}
\label{fig:qcorr}
\end{figure}
}

\def\figureJ{
\begin{figure}
\centerline{\epsfig{file=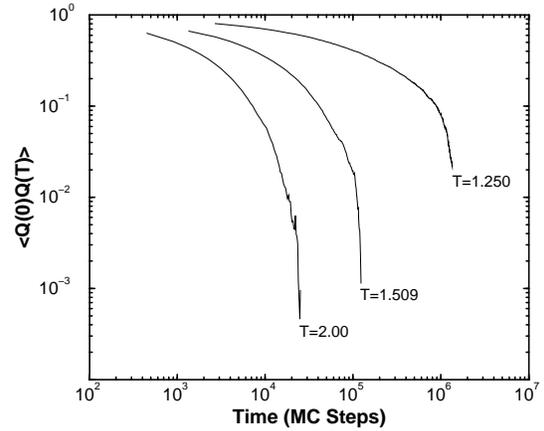,width=\figurewidth}}
\caption{$Q$ Autocorrelation function plotted on a log-log scale for
  several temperatures. As the temperature decreases the correlation
  time increases.}
\label{fig:qcorr1}
\end{figure}
}

\def\figureK{
\begin{figure}
  \centerline{\epsfig{file=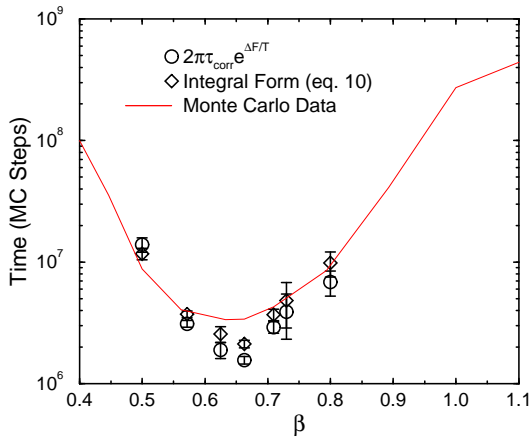,width=\figurewidth}}
\caption{Comparison of mean first passage times from Monte Carlo
  simulations with folding times calculated both the discrete from of
  the double integral (eqs.~\protect\ref{eq:integralrate} and
  \protect\ref{eq:sum}) and the simple rate formula $2\pi\taucorr
  e^{\beta\Delta F}$ which assumes a constant diffusion coefficient
  and that the curvature at the top and bottom of the free energy are
  the same.}
\label{fig:kramers}
\end{figure}
}

\def\figureL{
\begin{figure}
  \centerline{\epsfig{file=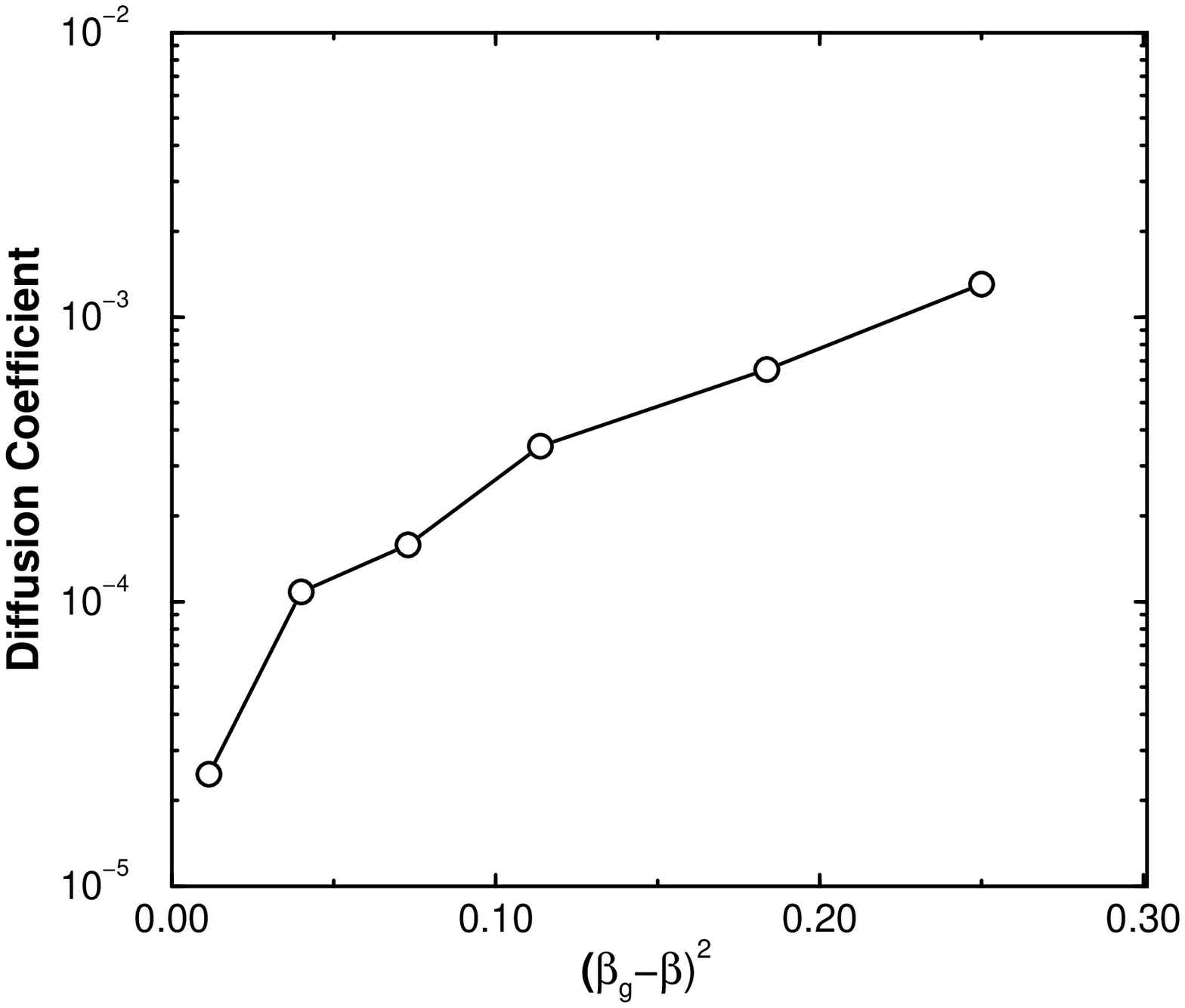,width=\figurewidth}}
\caption{Diffusion coefficient plotted versus $(\beta_g - \beta)^2$}
\label{fig:diff2}
\end{figure}
}

\def\figureM{
\begin{figure}
  \centerline{\epsfig{file=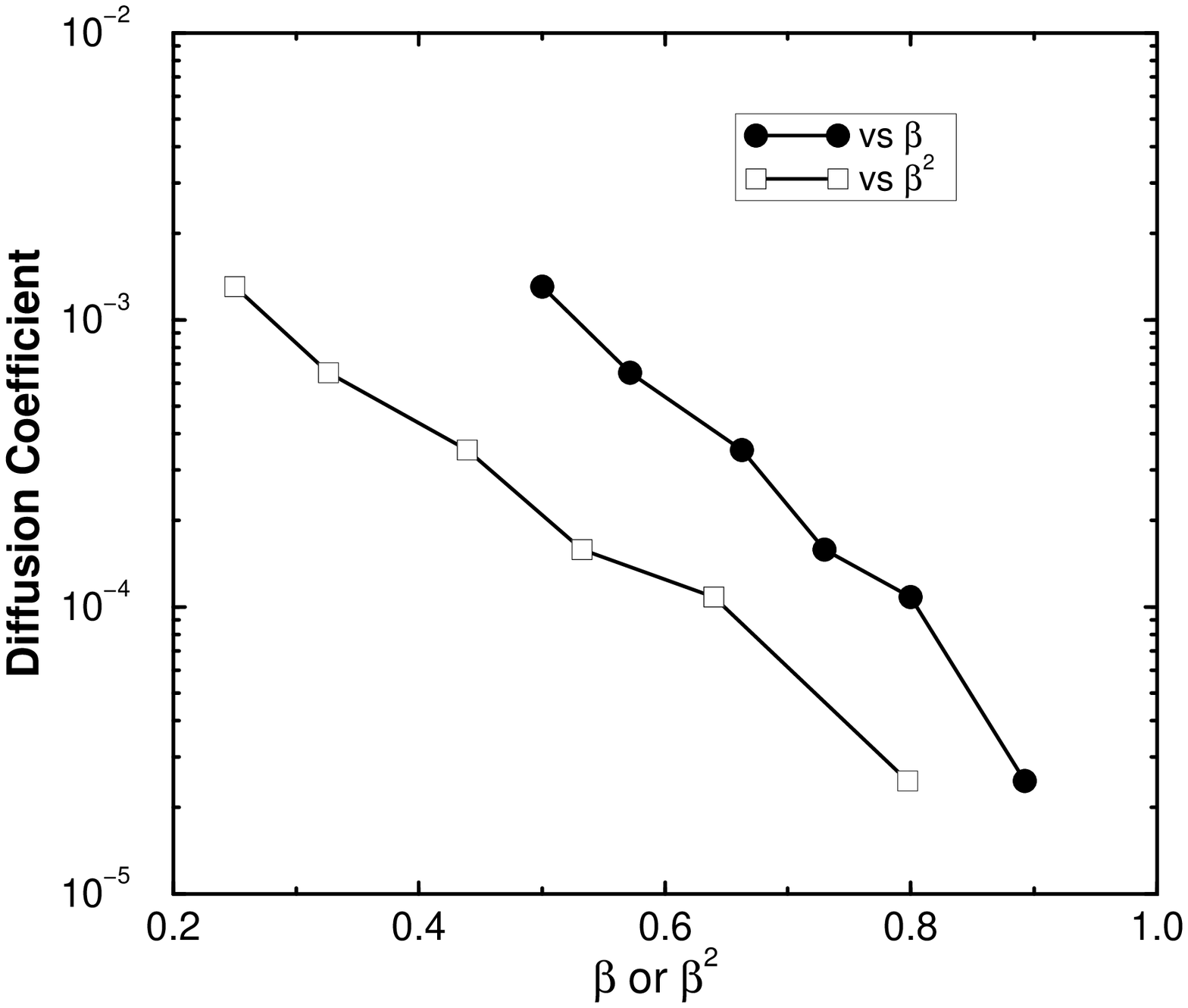,width=\figurewidth}}
\caption{Diffusion coefficient plotted versus $\beta$ and $\beta^2$}
\label{fig:diff}
\end{figure}
}


\begin{abstract}
  The quantitative description of model protein folding kinetics using
  a diffusive collective reaction coordinate is examined.  Direct
  folding kinetics, diffusional coefficients and free energy profiles
  are determined from Monte Carlo simulations of a 27-mer, 3 letter
  code lattice model, which corresponds roughly to a small helical
  protein.  Analytic folding calculations, using simple diffusive rate
  theory, agree extremely well with the full simulation results.
  Folding in this system is best seen as a diffusive, funnel-like
  process.

\end{abstract}

\section{Introduction} 

Protein folding is a collective self-organization process,
conventionally described as a chemical reaction. However, this process
generally does not occur by an obligate series of discrete
intermediates, a ``pathway,'' but by a multiplicity of routes down a
folding funnel.%
\cite{Leopold92,Bryngelson95,Onuchic95,Wolynes95,Dill93,Dill95,Boczko95}
Dynamics within a folding funnel involves the progressive formation of
an ensemble of partially ordered structures, which is transiently
impeded in its flow toward the folded structure by trapping in local
minima on the energy landscape. As one proceeds down the folding
funnel, the different ensembles of partially ordered structures can be
conveniently described by one or more collective reaction coordinates
or order parameters. Thermodynamically this funnel is characterized by
a free energy that is a function of the reaction coordinate which is
determined by the competition between the energy and entropy.  A
crucial feature of the funnel description is the concerted change in
both the energy and the entropy as one moves along the reaction
coordinate. As the entropy decreases so does the energy.  The gradient
of the free energy determines the average drift up or down the funnel.
Superimposed on this drift is a stochastic motion whose statistics
depends on the jumps between local minima. To first approximation this
process can be described as diffusion. Folding rates are determined
both by the free energy profile of the funnel and the stochastic
dynamics of the reaction coordinates. In this paper we study the
dynamics of a reaction coordinate describing the folding funnel of a
lattice model with thermodynamic parameters which theoretical
arguments suggest realistically correspond with fast folding small
helical proteins.\cite{Onuchic95} We determine from simulation both
the free energy profiles and effective configurational diffusion
coefficients for a folding reaction coordinate as a function of
temperature.  Folding times, also determined by simulations, are
compared to analytical calculations based on these quantities.

The kinetic analysis of a single folding funnel is most appropriate
for the fast folding proteins that mainly flow downhill progressively
in energy to the folded state. During folding, even for this case,
trapping in local minima will occur for times very short compared the
overall folding time. Alternatively, when the residence time in these
traps becomes too long leading to a substantial slow down of the
folding time, the traps can be thought of as creating additional
funnels. This occurs near the glass transition of the heteropolymer.
Good folding sequences will fold at a temperature $\Tf$, which is
above the glass transition temperature, $\Tg$, and have a single
dominant folding funnel.  As the chain moves downhill energetically in
its dominant funnel and becomes more similar to the native structure ,
the configurational entropy of the chain (number of available states)
is reduced. For the model discussed here, a single order parameter
suffices to measure similarity. For real proteins more order
parameters may be necessary to quantify the similarity of the
configuration to the native structure for example: degree of collapse,
helicity as well as fraction of correct contacts and dihedral angles
in the general case.  A free energy surface as a function of these
order parameters can then be calculated. For the case of a single
dominant funnel, these order parameters may also be associated with
reaction coordinates. The motion of these reaction coordinates will be
largely diffusive due to the transient trapping.  For the single
funnel case, the folding time is generally determined both by the
difficulty to overcome the free energy barrier and a prefactor that
depends on the ruggedness of the landscape that enters via the
diffusion coefficients.  This general quantitative description of
folding using diffusive coordinates along with energy landscape ideas
to account for trapping was introduced by Bryngelson and
Wolynes\cite{Bryngelson87,Bryngelson89} (BW). A short overview of
their ideas is given in the next section. Many of the qualitative
ideas from that description (e.g. the importance of the relationship
between $\Tf$ and $\Tg$ for kinetics) have been qualitatively
confirmed in lattice studies of model proteins\cite{Socci94b,Socci95}
and others have studied the dynamical properties of these model
systems.\cite{Miller92,Chan93a,Chan94,Sali94,Dill95}  The main goal
of the present paper is to quantitatively compare the BW expressions
for the folding time and effective diffusion coefficients with the
simulation results over a range of thermodynamics conditions for a
model that theory suggests realistically corresponds with the faster
folding small proteins.

In a previous paper,\cite{Onuchic95} we started this analysis.  There
it was shown that at the folding temperature ($\Tf$) the trajectories
of appropriate collective reaction coordinates are Brownian and must
surmount a modest thermodynamic free energy barrier.  The folding time
at $\Tf$ was well described by the diffusive rate formula. This work
shows how, using a combination of simulation results and analytical
theories, to formulate a law of corresponding states relating simple
lattice models to small helical proteins. Since free energy surfaces
are temperature dependent, various scenarios for the folding mechanism
apply at different temperatures.\cite{Bryngelson95}  By testing the
validity of analytical formulas over a wide temperature range, we hope
to provide a route to use theoretical descriptions to design and
understand quantitative experiments.


\section{A summary of the Bryngelson and Wolynes Energy Landscape Picture}

The energy landscape of any heteropolymer is complex. Because of the
possibility of making many contacts involving residues that are
distant in sequence, the energy landscape is rough. This is an effect
of frustration. Protein-like heteropolymers obey a principle of
minimal frustration, leading to an additional funnel-like aspect of
the landscape. To quantify this, we recognize that the energy
landscape is stratified, that is to say the statistical characteristics
of the landscape depend on the distance from, or equivalently the
similarity to, the ideal native structure. This similarity measure is
known as an order parameter in the physics of phase transitions, and
for small systems such as proteins can also be used as a reaction
coordinate for computing the folding rate. (Here this the reaction
coordinate will be called $n$.) Clearly such a reaction coordinate is not
unique. A picture of such a stratified landscape with kinetic
connections is shown in figure~\ref{fig:landscape}. 
\figureA
BW show that to a
first approximation, the folding time can be computed by first
grouping together states with in a stratum having a common value of
the reaction coordinate. A diffusion equation for the probability flow
between strata is derived under the assumption that the reaction
coordinate can only change by relatively small steps, and it is
written as
\begin{eqnarray}
\lefteqn{{\partial P (n,t) \over \partial t} = }\nonumber \\
&& { \partial \over \partial n}
\left\{ D(n) \left[ { \partial  P (n,t) \over \partial n}
+  P (n,t)  { \partial\  \beta F(n) \over \partial n} \right] \right\}\qquad
\end{eqnarray}
The average direction of the flow is given by the gradient of the free
energy. The diffusion coefficient, $D(n)$, depends on the trapping in
local minima and reflects the ruggedness of the energy landscape in 
the system proximity of the glass transition.

The free energy $F(n)$ incorporates the balance between two terms, the
energy that is decreased as the native state is approached and the
entropic term $-T S(n)$ which decreases with unfolding. The shape of
the free energy profile is therefore strongly temperature dependent as
illustrated in figure~1b. At high temperatures, folding is an uphill
process thermodynamically so folding is exponentially suppressed. At
the folding temperature, the free energy profile is typically bistable
with a small thermodynamic barrier which arises from the incomplete
cancellation of the entropy by the energy as the systems descends
through the funnel. At low temperatures, folding becomes a downhill
process. Thus, if the diffusion coefficient were temperature
independent, the rate of folding would have a dependence on the
thermodynamic driving force (which depends on $1/T$) just as a solid
state rectifier depends on the applied voltage (as in the famous
example of Feynman). The folding time can be written as a double
integral

\begin{equation}
\tauf = \int_{n_{\rm unf}}^{n_{\rm fold}} dn \int_0^n d n' { \exp \left[
    \beta F(n) - \beta F(n') \right] \over D(n) }
\label{eq:integralrate}
\end{equation}

When the barrier exists, as at $T_f$, $F(n)$ has a double-well with
the bottom of one well close to $n_{\rm unf}$ and the bottom of the other
well at $n_{\rm fold}$. In this situation this double integral can be
approximated by a Kramer's like law,\cite{Kramers40,Onuchic88} 
\begin{equation}
\tauf = \left( { 2 \pi \over \beta} \right)^{1/2} 
\frac{\exp\left\{\beta\left[\bar F(n_{t})-F(n_{\rm unf})\right]\right\}}%
{D_0 \ \omega_{\rm unf} \ \bar \omega_{\rm fold}}
\label{eq:kramersrate}
\end{equation}
where 
\begin{equation}
\bar F (n) = F(n) - T X(n)
\end{equation}
and
\begin{equation}
X(n) = \log \left[ {D(n) \over D_0} \right],
\end{equation}
and $\omega_{\rm unf}$ is the curvature around $n_{\rm unf}$ and $
\bar \omega_{\rm fold}$ is the curvature in the top of the barrier.
$D_0$ is the effective diffusion coefficient on an equivalent flat
landscape.

The prefactor of Kramer's expression reflects the multiple recrossings
of the barrier through diffusion that is controlled by trapping. When
the process is entirely downhill, the double integral becomes simply
proportional to the diffusion coefficient, and only weakly depends on
the slope of the free energy gradient.  

The configurational diffusion coefficient depends both on the local
moves allowed to the protein and the energy landscape topography. BW
analyzes configurational diffusion by assuming the limit of an
uncorrelated rugged energy landscape and Metropolis rules. This is, of
course, a caricature of the dynamics in which the landscape will have
correlations and the barriers to escape from traps may be surmounted
by successively melting out local clusters rather than globally
changing the protein conformation.  According to the BW analysis, the
diffusion coefficient has a strong temperature dependence that arises
from the necessity to escape from local minima on the rough energy
landscape. At high temperatures, $D$ follows a Ferry law typical of
glasses,

%
\begin{equation}
D(T,n) = D_0 \ \exp[-\beta^2 \Delta E^2(n)]
\end{equation}
where $\Delta E^2(n)$ is the local mean square fluctuation in
energy. At intermediate temperatures, the strongly non-Arrhenius
temperature dependence becomes itself moderated
\begin{eqnarray}
\lefteqn{D(T,n)=}\nonumber\\
&&D_0\exp\left\{-S^*(n) + [\beta_g(n) -\beta]^2  \Delta E^2(n)
\right\}\quad
\end{eqnarray}
This equation is valid for temperatures between $\Tg(n)$ and
$2\Tg(n)$. This form arises because of the assumed Metropolis
dynamics.  The characteristic temperature is the ideal glass
transition where the configurational entropy would vanish at
equilibrium. This is given by
the condition
\begin{equation}
\Tg(n)= {  \Delta E(n) \over [2 S^*(n)]^{1/2} },
\end{equation}
where $S^*(n)$ is configurational entropy at $n$.  We see that at
$\Tg$, the diffusion coefficient is diminished from its bare value by
a factor of the total number configuration states, thus giving rise to
a possibility of a Levinthal paradox. Below $\Tg$, we expect both that
the diffusion coefficient will still possess an activation energy but
the activation energy will not be self-averaging and that the free
energy itself can fluctuate considerably and will be very sensitive to
details of the model and to the sequence of the heteropolymer. In the
theory of disordered systems this is known as the emergence of non
self-averaging behavior, the hallmark of the ideal thermodynamic spin
glass transition in mesoscopic systems. Below $\Tg$, the slow folding
process can be better described by a few kinetic pathways than by the
statistical average laws appropriate for the faster events. Both the
temperature dependence of $D$ and the interplay of 
entropy and energetic ruggedness in the driving force lead to a
parabolic dependence of the folding rate on the inverse temperature as
shown in figure~\ref{fig:parabola}. 
\figureB
This rough form has been confirmed
in many lattice simulations. Leite and Onuchic\cite{Leite95}
have shown that for an energy landscape model for solvent
polarization, analogous to the BW landscape, there is a gradual
transition into a slow non-self averaging dynamics. Fluctuations
gradually dominates the kinetics below this temperature.

Folding rates depends on both diffusion and free energy profile in a
way that is significantly different from the standard transition state
theory (TST) The extent to which the rate is determined by the free
energy profile versus the diffusion coefficient, that incorporates the
multiple barrier crossing, depends on the choice for the reaction
coordinate. Notice however, that while it may be impossible or at least
difficult to find a reaction coordinate for which TST is exact, the
diffusion formula (eq.~2) is essentially invariant to this choice as
long as the elementary moves are reasonably local for this coordinate.

In the original BW treatment, this reaction coordinate was represented
by an order parameter $\rho$ that measured the similarity between any
given configuration and the native state of the protein in terms of
the fraction of correct dihedral angles, a coordinate which is
manifestly local because of the elementary moves of the protein at the
microscopic level are controlled by isomerization of the peptide bond.
Another possible reaction coordinate is the number of correct
contacts, a coordinate that is only partially local.



\section{Configurational diffusion for lattice models in proteins}

Are these general ideas developed by Bryngelson and Wolynes (BW) valid
for quantitatively predicting folding times in model proteins with a
realistic energy landscape topography? We show in this section that as
long as the glass transition falls after the transition region (top of
the barrier in the free energy profile for the collective reaction
coordinate) that this is the case.  In this limit the single dominant
funnel picture is appropriate.  The system under study is the designed
three--letter code 27-mer lattice model (see figure~\ref{fig:cube})
used in our recent studies.\cite{Socci94b,Socci95,Onuchic95}  
\figureC
Although this 27-mer is far from being a real protein, it has been
shown\cite{Onuchic95} that by developing a law of corresponding
states, the results obtained for such a system can be used to describe
a small $\alpha$-helical protein. In these simulations the units of
temperature and energy are such that $k_b=1$. This still leaves an
arbitrary scale factor since the only important quantity is the ratio
of energy to temperature. We have chosen small, integer values for the
coupling energies for convenience and efficiency. The kinetic glass
temperature is $T_g \sim 1$.

\figureD

Figure~\ref{fig:traj} shows the time evolution at $T_f$ for energy and
the $Q$ order parameter. This $Q$ parameter is a measure of similarity
to the native state and is equal to the number of {\em correct}
contacts (i.e. contacts that exist in the native state). It ranges
from 0, no correct contacts, to 28, the maximum number of possible
contacts.  Figure~\ref{fig:freeq} plots the energy, entropy and free
energy as a function of $Q$ for various temperatures. At temperatures
above the glass temperature, there is a significant gradient in both
energy and entropy.
\figureE
These profiles are computed from the density of states obtained using
the Monte Carlo histogram technique.\cite{Socci95} Starting from a
random configuration, collapse occurs at times very short compared to
the folding time. Thus in this parameter range, the radius of gyration
need not be considered as a separate dynamical reaction coordinate;
however, in determining the free energies one must note that the mean
radius of gyration does vary with temperature. A molten globule band,
where configurations have an average of 20 contacts and $Q=7.5$
($\approx27\%$ similarity to the native state), describes the region
where this sequence spends most of its time on its way into the folded
state. The shape of the free energy in the vicinity of this mean $Q$
is quasi- harmonic, although in fact large deviations are possible at
high temperatures. As shown in our earlier work, since at $\Tf$ the
local glass transition as a function of $Q$ occurs after the
transition region has been overcome, the folding time is determined
primarily by the time taken to overcome the free energy barrier, i.e.,
to cross the transition region. Figure~\ref{fig:trajfq} shows a
folding trajectory
of the $Q$ coordinate superimposed on a plot of the free energy. 
\figureF
The plot shows a time span which is roughly one quarter of the mean
folding time ($\tauf$) at this temperature. Most of the trajectory
consists of diffusive motion about the molten globule region. Once the
barrier has been surmounted folding occurs rapidly, taking roughly
$10^5$ Monte Carlo steps ($\approx.03\tauf$).

However, to estimate the folding times at a variety of temperatures,
knowledge of the free energy barrier alone is not sufficient.
Information about the dynamics must be obtained by calculating the
configurational diffusion coefficient through the complex energy
landscape.  As described in the previous section, when a single
reaction coordinate is considered, for example $Q$, $\tauf$ can be
computed using the double integral give by eq.~2. In general the
diffusion coefficient will depend on $Q$ but, one more simplification
is assumed here. Only the average value of $D$, computed for states in
the molten globule band, is inferred from simulations. We do this by
computing the correlation function of the fluctuations of the reaction
coordinate $\Delta Q(t)$. Within the quasi-harmonic diffusive
approximation this correlation function should decay exponentially at
long times. The configurational diffusion coefficient will be related
to the corresponding correlation time and the mean square
instantaneous value of the reaction coordinate fluctuations, $\Delta
Q^2(T)$. (This calculation is constrained for values of $Q$ before the
transition region. ) At a given temperature, the diffusive harmonic
model gives $D = \Delta Q^2(T) / \taucorr(T)$. Since $\taucorr <
\tauf$, it is a quantity that is much easier to obtain from numerical
simulations for more complex systems such as atomistic simulations of
proteins. For experimentalists we especially wish to point out this
viewpoint and the corresponding approximation allow one to use
dynamic probes of fluctuations in the molten globule state at
equilibrium to predict the rate of the non-equilibrium folding
process.

In figure~\ref{fig:tcorr} we show the simulated $\taucorr$ and
diffusion coefficient for various temperatures above the glass
transition. 
\figureG
The autocorrelation time as been calculated in the molten
globule region. To do this, the correlation function $<Q(t)Q(0)>$ was
calculated over paths which were confined to the molten globule
region of the conformation space. For these calculations we define any
conformation with $Q<17$ as being in the molten globule region. The
decays are in fact non-exponential for short times. We have not
analyzed the non-exponentiality in detail. The information would
provide the dynamics of individual conformation steps and should
contain information on the distribution of substate lifetimes. We
identify the long time decay with the effective diffusive harmonic
value.  A more complete model should account for this short time decay
via a frequency dependent diffusion coefficient as proposed by BW.
Frequency effects may be important for determining the prefactor of
barrier crossing as discussed by Grote and Hynes for simple chemical
reactions.\cite{Grote80} 


\section{Details of simulations and comparison with BW analysis}


To study the dynamics of this model we used the Monte Carlo algorithm
with local moves, which include end moves, corner flips and $90^\circ$
crankshaft moves. The excluded volume constraint was enforced by not
allowing any multiple occupancy. (Further details can be found in a
previous work.\cite{Socci94b}) To calculate the mean first passage
time ($\tauf$) as a function of temperature the following procedure
was used. For each temperature many simulations were performed, each
starting from a different random unfolded initial condition. The
simulations were run until the chain found the folded state or until a
maximum number of time steps $\taumax$ had elapsed.  The mean first
passage time is simply the average of the folding times for the
different runs. For an intermediate range of temperatures all of the
runs would find the folded state within $\taumax$ steps. However for
very low and high temperatures only some fraction would find the
folded state in the allowed time. For these temperatures the mean
calculated is a {\em lower bound} to the actual mean first passage
time. This is because for trials that do not fold within $\taumax$
steps, we average in $\taumax$ which is less than the actually folding
time for that run.  This is done for practical reasons since we have a
finite amount of computer time for the simulations. The key point is
that $\taumax$ is much greater than $\tauf$ for a broad range of
temperatures; which in fact, it is. Figure~\ref{fig:tauf} shows the
mean first passage time simulated for a range of temperatures.

At low temperatures, $\tauf$ grows rapidly. This slow down is caused
by trapping in local minima. It is well known that heteropolymers
undergo an ideal glass transition at low enough temperature. The
random energy model estimate of this temperature for the present
system is $\Tg^{REM}= \Delta E^2 / \sqrt{2 S_L}$.  The measured
energetic fluctuations $\Delta E^2$'s are $51$ at $\tauf$ and $22$ at
$T=1.1$ (slightly above the glass transition). The respective
configurational entropies for the collapsed states, $S_L$, are around
20 and 12. Both cases provide a $\Tg^{REM} \sim 1.0$. This agrees with
the critical temperature at which replica symmetry breaking sets in.
This is determined by finding the temperature at which $Y = \sum_i
P_i^2$ becomes of order 1 (see figure~\ref{fig:yq}).
\figureH
$P_i$ is the Boltzmann occupation of a microstate. $Y$ is not the
perfect measure of replica symmetry breaking since it assumes the
basin of attraction of a microstate can be identified with the state
itself. The corresponding neglect of correlation between these high
$Q$ states, however, is able to provide a reasonable estimate of $\Tg$
for a finite mesoscopic system.

These estimates of $\Tg$ are very similar to those of the {\em kinetic
  glass transition} which is the temperature at which the folding time
slows down substantially. We define the kinetic glass transition
temperature ($\Tg$) to be the temperature at which the folding time is
sufficiently slower than the fastest folding time observed. There are
several ways of choosing this point which will give roughly the same
answer. We choose $\Tg$, to be the temperature at which
$\tauf=\frac{1}{2}\taumax$. For this sequence, $\Tg\approx1$.

In addition to the kinetic runs to calculate $\tauf$, we performed a
number of thermodynamic runs to determine the free energy profiles.
Again most of the details can be found in a previous
work.\cite{Socci95}   We used the Monte Carlo histogram technique
which allows us to calculate extensive quantities like the free
energy.  The basic idea is to store a histogram as a function of $Q$
and $E$.  These histograms measure the probability that a given
$(Q,E)$ pair will occur. From these we can calculate the density of
states $n(Q,E)$ up to a multiplicative constant (which can be
determined since we know the $n(1,E_{\rm min})=1$). We can then
calculate any thermodynamic quantity; in particular, we determined the
free energy as a function of $Q$ ($F[Q]$) for several temperatures.
The free energy is shown in figure~\ref{fig:freeq}. 
\figureI
For a broad range of temperatures the free energy has a rough double
well profile, with one minimum at the folded state ($Q=28$) and
another for the molten globule ($Q\approx5$--$7$). The transition
state varies from $\Qts=22$ at $T=2$ to $\Qts=16$ at the lower
temperatures. At first it might appear the transition state is at
$\Qts=25$ since this is the highest local maximum in the plot at most
temperatures.\cite{Sali94a}  However, this turns out not to be the
correct transition point.  One must be careful in interpreting free
energy plots of this type. There are Monte Carlo moves that connect
states with $Q=23$ directly to the ground state short circuiting this
barrier. In fact any barrier that is smaller in width than 5 can also
be by passed by certain Monte Carlo moves. Note, the correct barriers
from 22 to 16 are much broader. In fact it is better to think in terms
of a transition ``region'' rather than a unique transition point. The
way in which we identified this region was to look at the time
trajectories (figure~\ref{fig:traj}).  Since we are in a diffusive
regime, a trajectory that reaches the transition region should have
some probability of diffusing back down the barrier rather than
crossing over. For $\Qts$ values between 16--22 we see such an effect.
However, by the time $\Qts=25$ is chosen there would be virtually no
trajectories that cross back without first reaching the folded state.
We have also previously used a 2 dimensional analysis of the
trajectories and free energy surfaces and again we find the transition
region to be between 16-22 and not at 25.\cite{Onuchic95}  Since we
know that the transition state is somewhere between 16 and 22 we
define a specific transition point, for use in calculating the barrier
for the Kramer's rate equation, as the value of $Q$ that is a maximum
{\em in this region}.

In addition to the free energy barrier we need to measure the
diffusion coefficient in this system in order to calculate the folding
time. We compute the diffusion coefficient by measuring the
autocorrelation time of $Q$. Specifically we calculate:
\begin{equation}
C_{Q}(\Delta) = 
 \frac{\left\langle Q(t)Q(t+\Delta)\right\rangle-\left\langle Q(t)\right\rangle^2}%
{\left\langle Q^2(t)\right\rangle-\left\langle Q(t)\right\rangle^2}
\label{eq:autocorr}
\end{equation}
Figure~\ref{fig:qcorr} shows a semi-log plot of the auto correlation
function at $T=1.509$, the folding temperature. The figure is
multi-exponential with a short fast decay followed by a much longer,
slower decay. The final drop-off is mostly due to errors in sampling at
very long times. We are interested in the long time decay and
calculated the exponent associated with this by fitting a function of
several exponentials (usually 3 or 4) to this plot. We repeated this
calculation at several temperatures (note that we can not use the
histogram method here do the large amount of histogram information
that would have to be stored). Figure~\ref{fig:qcorr1} is a log-log
plot of the autocorrelation function for several temperatures. As the
temperature is decreased the autocorrelation time increases. 
\figureJ
Finally the diffusion coefficient was estimated as $D=\Delta
Q^2/\taucorr$.

%
%

With the diffusion coefficients and free energy surfaces in hand, it
is now possible to test the analytical prediction given by
eq.~\ref{eq:integralrate}.  Using the discrete form of the full
double integral (with a constant diffusion coefficient):
\begin{equation}
\tauf = \frac{1}{D_0}\sum_{Q=0}^{23} \sum_{Q'=0}^{Q-1}
                       \exp\left\{\beta\left(F[Q]-F[Q']\right)\right\}
\label{eq:sum}
\end{equation}
we obtain the results presented in figure~\ref{fig:kramers}. 
\figureK
Note the sum is not taken to $Q=28$ since we do not expect the
diffusion coefficient to remain constant in that region (between
$Q=23$--$28$).  However, the time it takes to go from $Q=23$ to the
folded state ($Q=28$) is small compared to $\tauf$ (as seen in
figure~\ref{fig:trajfq}) so the error from truncating the sum will
also be small. Overcoming the barrier ($Q\approx17$) is the limiting
step. Additionally, we can also use a simplified form of
eq.~\ref{eq:integralrate} which assumes that the free energy surface
is well approximated by a parabolic potential around the molten
globule states that extends all the way to the barrier
(eq.~\ref{eq:kramersrate}).  If we further assume that $D(n)=D_0$ and
that the two curvatures are the same then the folding time is given by
$\tauf = 2\pi\taucorr e^{\Delta F/T}$ (which is also shown in
figure~\ref{fig:kramers}).  This simple result is useful because it
depends only on the correlation time in the molten globule and the
free energy barrier.  The agreement with the full formula is
remarkable. For both formulas, the fastest time at the optimal kinetic
folding temperature is obtained to well within a factor of two.  Also,
an extremely good description of the behavior for the slow down
anticipated for low and high temperatures is obtained. Thus we see the
analytical theory based on the actual molten globule dynamics and
funnel free energy profile is not simply qualitatively correct but can
be used for quantitative predictions of the folding time over a wide
thermodynamic range.

Up to this point, the values for the diffusion coefficients have been
obtained directly from the simulation data for the time correlation
functions around the molten globule. It is interesting now to compare
the obtained results with the existing models of configurational
diffusion, particularly the BW theory.  Since our results fall in the
range between $\Tg$ and $2\Tg$, the intermediate temperature formula
given by eq.~7 would be seem to be the appropriate comparison.
Figure~\ref{fig:diff2} shows a plot of the diffusion coefficients as a
function of $(\beta_g - \beta)^2$. 
\figureL
The slope obtained from the
semi-log plot provides a fitted value for $\Delta E^2$ of $\sim
15$. This is a little bit smaller than the $\Delta E^2$'s obtained
directly from simulations at this temperature range.  This is expected
because correlations in the landscape cause only part of the total
energetic ruggedness to come into the activation barriers for escaping
from traps.  Qualitatively agreement between the model and simulations
suggests that while the $REM$ results are a first approximation, the
correlations in the energy landscape are significant and the effective
fluctuations for escape from a trap involves only a fraction of the
chain.\cite{Leite95,Wang95}  We would expect these correlation effects
to become more important as the effective chain length of the protein
grows, however we should bear in mind that the same qualitative
behavior still applies. In the temperature range studied, we note the
data could be equally well fitted by an Arrhenius temperature plot
(see figure~\ref{fig:diff}). 
\figureM
The effective activation energy is $\sim 10$ ($\sim 7 k_B T_f$).  This
very large number implies that the effective moves of $Q$ require
substantial organization of the protein since the stability gap is
only $20 k_B T_f$. The simple Ferry law fit is also adequate providing
a $\Delta E^2 \sim 8$ (see eq.~4 and figure~\ref{fig:diff}).

\section{Conclusions}






%

Protein folding is remarkable as a chemical reaction with a highly
non-Arrhenius temperature dependence. Although for real proteins some
of this behavior is doubtless due to the entropic nature of
hydrophobic forces, we see here that this unusual temperature
dependence also arises from the competition
between trapping in misfolded states and drift down the folding funnel
which is itself a competition between entropy and energy. These
effects can be described through collective coordinates, which make
quantitative the nature of the funnel. Our results establish that a
description in terms of a single collective coordinate suffices to
explain folding kinetics of small lattice models over a wide range of
temperatures. We believe that a similar set of concepts can be used
for the smaller real proteins because of their correspondence with the
lattice models. The diffusive dynamics should be contrasted to the
more traditional transition state theory. These simulations suggest
that if transition state theory is used a very complex reaction
coordinate, reflecting the detailed trapping dynamics, would have to
be constructed. The single collective diffusive coordinate picture on
the other hand is much more robust and can make use of experimental
information about the dynamics of fluctuations in the molten globule,
as well as simple thermodynamic measurements and theories about the
molten globule state.

The lattice model studied here are described quite well by a single
order parameter or reaction coordinate. Various scenarios for protein
folding require the introduction of a few more collective coordinates
to describe the funnel. Even the small helical proteins, that
corresponding in a thermodynamic sense with the 27-mer lattice,
require the introduction of coordinates describing the amount of
helicity, degree of collapse and liquid crystallinity of the dynamical
molten globules. If these order parameters are constant or their
dynamics is ``slaved'' to the tertiary ordering parameter the
one-dimensional diffusive theory for funnel dynamics will suffice for
real proteins just as in the model systems. If the collective
diffusion coefficients are wildly different for these extra degrees of
freedom or when there is a free energy profile for the coupled order
parameters with several minima, one must generalize the Kramer's
expression to a few more dimensions.  We believe such a
few-dimensional generalization of the current diffusive dynamical
description should suffice for the smaller fast folding protein
molecules. On the other hand, sufficiently large proteins doubtless
require a description in terms of geometrically localized order
parameters not just a globally defined coordinate. One example of
geometrically localized phenomena is provided by
foldons.\cite{Panchenko95}   Foldons are kinetically autonomous
folding subunits, that are expected to exist in the larger proteins.
Each foldon will have its own funnel. A related description involving
local collective coordinates necessary for describing critical nuclei
may also be helpful. Some arguments suggest critical nuclei may be
large\cite{Bryngelson90} so the system would be well treated by
global reaction coordinate while others suggest nuclei may
small\cite{Thirumalai95b} or indeed possibly
unique.\cite{Abkevich94b}  Any of these cases can be accommodated by
the diffusive funnel dynamics picture once the coordinates are
specified. We expect the diffusive funnel dynamics picture
(discussed here and previously by us\cite{Onuchic95}) will provide a
convenient quantitative framework to analyze both simulation and
laboratory experiments.

\par\medskip\par\hrule\par\bigskip\par

\end{document}